\documentclass[review]{elsarticle}
\usepackage[latin1]{inputenc}
\usepackage[T1]{fontenc}
\usepackage{amsmath, amsthm, amssymb}
\usepackage{array}

\newtheorem{definition}{Definition}[section]
\newtheorem{theorem}[definition]{Theorem}

\newtheorem{remark}[definition]{Remark}

\newtheorem{example}[definition]{Example}

\begin{document}

\title{Erasure decoding of convolutional codes using first order representations}


\author{Julia Lieb\corref{cor1}%
}
\ead{julia.lieb@math.uzh.ch}
\address{Institute of Mathematics, University of Zurich
Winterthurerstrasse 190, 8057 Zurich, Switzerland}


\author{Joachim Rosenthal%
}
\ead{rosenthal@math.uzh.ch}
\address{Institute of Mathematics, University of Zurich
Winterthurerstrasse 190, 8057 Zurich, Switzerland}

\cortext[cor1]{Corresponding author}

\setcounter{page}{1}

\pagenumbering{arabic}
\pagestyle{plain}

\begin{abstract}
In this paper, we employ the linear systems representation of a convolutional code to develop a decoding algorithm for convolutional codes over the erasure channel. We study the decoding problem using the state space description and this provides in a natural way additional information. With respect to previously known decoding algorithms, our new algorithm has the advantage that it is able to reduce the decoding delay as well as the computational effort in the erasure recovery process. We describe which properties a convolutional code should have in order to obtain a good decoding performance and illustrate it with an example.
\end{abstract}


\maketitle

\section{Introduction}

In modern communication, especially over the Internet, the erasure channel is widely used for data transmission. In this type of channel the receiver knows if an arrived symbol is correct, as each symbol either arrives correctly or is erased. For example over the Internet messages are transmitted using packets and each packet comes with a check sum. The receiver knows that a packet is correct when the check sum is correct. Otherwise a packet is corrupted or simply is lost during transmission. An especially suitable class of codes for transmission over an erasure channel is the class of convolutional codes \cite{lc}. It is known that convolutional codes are closely related to discrete-time linear systems over finite fields, in fact each convolutional code has a so-called input-state-output (ISO) representation via such a linear system \cite{be,bch}. This correspondence was also used in \cite{c1,c2,c3} to study concatenated convolutional codes. Moreover, the connection between linear systems and convolutional codes was investigated in a more general setup in \cite{zerz}, where multidimensional codes and systems over finite rings were considered.
\\
Hence, decoding of a convolutional code can be viewed as finding the trajectory (consisting of input and output) of the corresponding linear system that is in some sense closest to the received data. The underlying distance measure one uses to identify the closest trajectory (i.e. the closest codeword) depends on the kind of channel that is used for data transmission.
This decoding process can also be interpreted as minimizing a cost function attached to the corresponding linear system, which measures the distance of a received word to a codeword or the distance of a measured trajectory  to a possible trajectory, respectively.
For the Euclidean metric over the field of real numbers $\mathbb R$, this is nothing else than solving the classical LQ problem, i.e. minimizing the cost function $\sum_{i=0}^{N-1}||u_i-\hat{u}_i||^2+||y_i-\hat{y}_i||^2$, where $\hat{u}\in(\mathbb R^m)^N$ and $\hat{y}\in(\mathbb R^p)^N$ are received and one wants to find an input $u\in(\mathbb R^m)^N$ and corresponding output $y\in(\mathbb R^p)^N$ of the linear system such that this cost function is minimized.
This problem is relatively easy to solve and it is known how to approach it for quite some time; see e.g. \cite{kai}.

However, for the setting of classical coding theory, where usually the Hamming metric over finite fields is used, it turns out to be in general a hard problem to minimize the corresponding cost function $\sum_{i=0}^{N-1}wt(u_i-\hat{u}_i)+wt(y_i-\hat{y}_i)$ with $\hat{u},u\in(\mathbb F^m)^N$ and $\hat{y},y\in(\mathbb F^p)^N$ for some finite field $\mathbb F$. The methods used to solve the LQ problem cannot be applied since the Hamming metric is not induced by a positive definite scalar product. However, the problem becomes much easier for transmission over an erasure channel as done with convolutional codes in this paper. In this setting, one introduces an additional symbol $\ast$ that stands for an erasure and considers $\mathbb F\cup\{\ast\}$ as set of symbols for the decoding. The Hamming metric can easily be extended to this new symbol space and we are going to minimize the same cost function. The big advantage when decoding over an erasure channel is that we know that all received symbols, i.e. all symbols except $\ast$ in $\hat{u}$ and $\hat{v}$, are correct and we only have to find a way to replace the unknowns $\ast$ be the original values to bring the cost function to its minimal value, which equals the number of erasures. 
It depends on the number of erasures if unique decoding is possible or if one gets a list of possible codewords. In this paper, we focus on unique decoding, i.e. we present an erasure decoding algorithm that skips part of the sequence if there are too many erasures such that unique decoding is not possible. Our algorithm exploits the ISO representation of a convolutional code via linear systems to recover the erasures in the received sequence. With respect to other erasure decoding algorithms for convolutional codes that can be found in the literature \cite{vp,j}, our systems theoretic approach has the advantage that the computational effort as well as the decoding delay can be reduced.


The paper is structured as follows. 
In Section 2, we give the necessary background on convolutional codes. In Section 3, we explain the correspondence of time-discrete linear systems and convolutional codes. In Section 4, we present our decoding algorithm, describe which properties a convolutional code should have to perform well with our algorithm and illustrate it with an example. In Section 5, we describe the advantages of our algorithm and in Section 6, we conclude with some remarks.

\section{Convolutional codes}

In this section, we start with some basics on convolutional codes.

\begin{definition}
An $(n,k)$ convolutional code $\cal C$ is defined as an $\mathbb F[z]$-submodule of $\mathbb F[z]^n$ of rank $k$.
As $\mathbb F[z]$ is a principal ideal domain, every submodule is free and hence, there exists a full column rank polynomial matrix $G(z) \in \mathbb F[z]^{n \times k}$ whose columns constitute a basis of $\cal C$, i.e.
\begin{eqnarray*}
  {\cal C} &=&Im_{\mathbb F[z]}G(z) \\
   &=& \{G(z)u(z) \, | \, u(z) \in \mathbb F[z]^k\}.
\end{eqnarray*}
\end{definition}

Such a polynomial matrix $G$ is called a \textbf{generator matrix} of $\cal C$.
A basis of an $\mathbb F[z]$-submodule of $\mathbb F[z]^n$, and therfore also a generator matrix of a convolutional code, is not unique. If $G(z)$ and $\tilde G(z)$ in $\mathbb F[z]^{n \times k}$ are two generator matrices of $\cal C$, then one has $G(z)=\tilde{G}(z)U(z)$ for some unimodular matrix $U(z)\in\mathbb F[z]^{k\times k}$ (a unimodular matrix is a polynomial matrix with a polynomial inverse). 

Another important parameter of a convolutional code is its \textbf{degree} $\delta$, which is defined as the highest (polynomial) degree of the $k\times k$ minors of any generator matrix $G(z)$ of the code.
An $(n,k)$ convolutional code with degree $\delta$ is denoted as $(n,k,\delta)$ convolutional code.
If $\delta_1,...,\delta_k$ are the column degrees (i.e. the largest degrees of any entry of a fixed column) of $G(z)$, then one has that $\delta\leq \delta_1+...+\delta_k$. Moreover, there always exists a generator matrix of $\cal C$ such that $\delta= \delta_1+...+\delta_k$ and we call such a generator matrix $\textbf{column reduced}$.

Furthermore, for the use over an erasure channel, it is a crucial property of a convolutional code to be \textbf{non-catastrophic}. A convolutional code is said to be non-catastrophic if one (and therefore each) of its generator matrices is right prime, i.e. if it admits a polynomial left inverse. The following theorem shows, why this property is so important.


\begin{theorem}\label{H}
Let $\cal C$ be an $(n,k)$ convolutional code. Then $\cal C$ is noncatastrophic if and only if there exists a so-called \textbf{parity-check matrix} for $\cal C$, i.e. a full row rank polynomial matrix $H(z) \in \mathbb F[z]^{(n-k) \times n}$ such that
\begin{eqnarray*}
  {\cal C} &=& Ker_{\mathbb F[z]}H(z) \\
   &=& \{v(z) \in \mathbb F[z]^n \, | \, H(z)v(z)=0\}.
\end{eqnarray*}
\end{theorem}


Parity-check matrices are common to be used for decoding of convolutional codes over the erasure channel. Recall hat, when transmitting over this kind of channel, each symbol is either received correctly or is not received at all. The first decoding algorithm of convolutional codes over the erasure channel using parity-check matrices can be found in \cite{vp}, variations of it in \cite{j} or \cite{2d}.
To investigate the capability of error correction of convolutional codes, it is necessary to define distance measures for these codes.

Therefore, we denote by the \textbf{Hamming weight} $wt(v)$ of $v\in\mathbb F^n$ the number of its nonzero components.
For $v(z)\in\mathbb F[z]^n$ with $\deg(v(z))=r$, we write $v(z)=v_r+\cdots+v_{0}z^{r}$ with $v_t\in\mathbb F^n$ for $t=0,\hdots,r$ and set $v_t=0\in\mathbb F^n$ for $t\not\in\{0,\hdots,r\}$. 
For $j\in\mathbb N_0$, we define the \textbf{j-th column distance} of a convolutional code $\mathcal{C}$ as
$$d_j^c(\mathcal{C}):=\min_{v(z)\in\mathcal{C}}\left\{\sum_{t=0}^j wt(v_{r-t})\ |\ v_r\neq 0\right\}.$$

The erasure correcting capability of a convolutional code increases with its column distances, which are upper bounded as the following theorem shows.

\begin{theorem}\cite{strongly}\label{ub}
Let $\mathcal{C}$ be an $(n,k,\delta)$ convolutional code. Then, it holds:
$$d_j^c (\mathcal{C}) \leq (n-k)(j + 1) + 1\ \ \text{for}\ \ j\in\mathbb N_0.$$
\end{theorem}

It is well-known that the column distances of a convolutional code could reach this upper bound only up to $j=L:=\left\lfloor\frac{\delta}{k}\right\rfloor+\left\lfloor\frac{\delta}{n-k}\right\rfloor$.

\begin{definition}\cite{mdp}
An $(n,k,\delta)$ convolutional code $\mathcal{C}$ is said to be
  \textbf{maximum distance profile (MDP)} if
$$d_j^c(\mathcal{C})=(n-k)(j+1)+1\quad \text{for}\ j=0,\hdots,L:=\left\lfloor\frac{\delta}{k}\right\rfloor+\left\lfloor\frac{\delta}{n-k}\right\rfloor$$
\end{definition}

If one has equality for some $j_0\in\mathbb N$ in Theorem  \ref{ub}, then one also has equality for $j\leq j_0$, see \cite{strongly}. Hence, it is sufficient to have equality for $j=L$ to obtain an MDP convolutional code. The following theorem presents criteria to check if a convolutional code is MDP.


\begin{theorem}\label{cd}\cite{strongly}
Let $\mathcal{C}$ have a column reduced generator matrix $G(z)=\sum_{i=0}^{\mu}G_iz^i\in\mathbb F[z]^{n\times k}$ and parity-check matrix $H(z)=\sum_{i=0}^{\nu}H_iz^i\in\mathbb F[z]^{(n-k)\times n}$. The following statements are equivalent:
\begin{itemize}
\item[(i)] $d_j^c (\mathcal{C})=(n-k)(j + 1) + 1$ 
\item[(ii)] $G^c_j:=\left[\begin{array}{ccc} G_0 & & 0\\ \vdots & \ddots &  \\ G_j & \hdots & G_0 \end{array}\right]$ where $G_i\equiv 0$ for $i>\mu$ has the property that every full size minor that is not trivially zero, i.e. zero for all choices of $G_1,\hdots,G_j$, is nonzero.
\item[(iii)] $H_j^c:=\left[\begin{array}{ccc} H_0 & & 0\\ \vdots & \ddots &  \\ H_j & \hdots & H_0 \end{array}\right]$ with $H_i\equiv 0$ for $i>\nu$ has the property that every full size minor that is not trivially zero
is nonzero.
\end{itemize}
\end{theorem}

The erasure decoding capability of an MDP convolutional code is stated in the following theorem.

\begin{theorem}\cite{vp}\label{mdpp}\ \\
If for an $(n,k,\delta)$ MDP convolutional code $\mathcal{C}$, in any sliding window of length at most $(L+1)n$ at most $(L+1)(n-k)$ erasures occur, then full error correction from left to right is possible.
\end{theorem}
%

\section{The linear systems representation of a convolutional code}

In this section, we consider discrete-time linear systems of the form
\begin{align}\label{eq:linsys}
x(\tau +1)&=Ax(\tau)+Bu(\tau) \nonumber\\
y(\tau)&=Cx(\tau)+Du(\tau)
\end{align}
with $A\in \mathbb{F}^{s\times s}, B\in
 \mathbb{F}^{s\times k}, C\in\mathbb F^{(n-k)\times s}, D\in\mathbb F^{(n-k)\times k}$, input $u\in\mathbb F^k$, state vector $x\in\mathbb F^s$, output $y\in\mathbb F^{n-k}$ and $s, \tau\in\mathbb N_0$.
We identify this system with the matrix-quadruple $(A,B,C,D)$. The function $T(z)=C(zI-A)^{-1}B+D$ is called \textbf{transfer function} of the linear system.

\begin{definition}\label{rom}\ \\
A linear system (\ref{eq:linsys}) is called 
\begin{itemize}
\item[(a)]
\textbf{reachable} if for each
$\xi \in \mathbb{F}^s$ there exist $\tau_{\ast}\in\mathbb N_0$ and a sequence of
inputs $u(0), \ldots, u(\tau_*) \in \mathbb{F}^k$ such that the sequence of
states $0=x(0), x(1), \ldots ,$ $ x(\tau_* +1)$ generated by
(\ref{eq:linsys}) satisfies $x(\tau_* +1)=\xi$. 
\item[(b)]
\textbf{observable} if $Cx(\tau)+Du(\tau)=C\tilde{x}(\tau)+Du(\tau)$ for all $\tau\in\mathbb N_0$ implies $x(\tau)=\tilde{x}(\tau)$ for all $\tau\in\mathbb N_0$. This means that the knowledge of the input and output sequences is sufficient to determine the sequence of states.
\item[(c)]
\textbf{minimal} if it is reachable and observable.
\end{itemize}
\end{definition}

Recall the following well-known characterization of
reachability and observability.

\begin{theorem}(Kalman test)\label{kal}\ \\
A linear system (\ref{eq:linsys}) is reachable if and only if the reachability matrix\\ $\mathcal{R}(A,B):=(B,AB,\ldots, A^{s-1}B)\in\mathbb F^{s\times sk}$ satisfies $\operatorname{rk}(\mathcal{R}(A,B))=s$ and observable if and only if the observability matrix $\mathcal{O}(A,C)=\begin{pmatrix}C\\ \vdots\\ CA^{s-1} \end{pmatrix}\in\mathbb F^{(n-k)s\times s}$
satiesfies $\operatorname{rk}(\mathcal{O}(A,B))=s$.

\end{theorem}

Next, we will explain how one can obtain a convolutional code from a linear system; see \cite{bch}. First, for $(A,B,C,D)\in\mathbb F^{s\times s}\times\mathbb F^{s\times k}\times\mathbb F^{(n-k)\times s}\times\mathbb F^{(n-k)\times k}$, we set $$H(z):=\left[\begin{array}{ccc}
zI-A & 0_{s\times(n-k)} & -B \\ 
-C & I_{n-k} & -D
\end{array}\right].$$
The set of $v(z)=\begin{pmatrix} y(z)\\u(z)\end{pmatrix}\in\mathbb F[z]^n$ with $y(z)\in\mathbb F[z]^{n-k}$ and $u(z)\in\mathbb F[z]^{k}$ for which there exists $x(z)\in\mathbb F[z]^s$ with
$H(z)\cdot[x(z)\ y(z)\ u(z)]^{\top}=0$
forms a submodule of $\mathbb F[z]^n$ of rank $k$ and thus, an $(n,k)$ convolutional code, denoted by $\mathcal{C}(A,B,C,D)$.

Moreover, if one writes $x(z)=x_0z^{\gamma}+\cdots+x_{\gamma}$,  $y(z)=y_0z^{\gamma}+\cdots+y_{\gamma}$ and $u(z)=u_0z^{\gamma}+\cdots+u_{\gamma}$ with $\gamma=\max(\deg(x),\deg(y),\deg(u))$, it holds
\begin{align*}
x_{\tau +1}&=Ax_{\tau}+Bu_{\tau} \\
y_{\tau}&=Cx_{\tau}+Du_{\tau}\\
(x_{\tau}, y_{\tau}, u_{\tau})&=0\ \text{for}\ \tau>\gamma.
\end{align*}
Furthermore, there exist $X\in\mathbb F[z]^{s\times k}, Y\in\mathbb F[z]^{(n-k)\times k}, U\in\mathbb F[z]^{k\times k}$ such that $\operatorname{ker}(H(z))=\operatorname{im}[X(z)^{\top}\ Y(z)^{\top}\ U(z)^{\top}]^{\top}$ and $G(z)=\begin{pmatrix}
Y(z)\\U(z)\end{pmatrix}$ is a generator matrix for $\mathcal{C}$ with $C(zI-A)^{-1}B+D=Y(z)U(z)^{-1}$, i.e. one is able to obtain a factorization of the transfer function of the linear system via the generator matrix of the corresponding convolutional code, and in the case that this convolutional code is non-catastrophic, one even obtains a coprime factorization of the transfer function.\\
On the other hand, for each $(n,k,\delta)$ convolutional code $\mathcal{C}$, there exists $(A,B,C,D)\in\mathbb F^{s\times s}\times\mathbb F^{s\times k}\times\mathbb F^{(n-k)\times s}\times\mathbb F^{(n-k)\times k}$ with $s\geq\delta$ such that $\mathcal{C}=\mathcal{C}(A,B,C,D)$. In this case, $(A,B,C,D)$ is called linear systems representation or input-state-output (ISO) representation of $\mathcal{C}$.
Besides, one can always choose $s=\delta$. In this case, $(A,B,C,D)$ is called a \textbf{minimal representation} of $\mathcal{C}$.

\begin{remark}
\normalfont
In the coding literature state space descriptions were often done in a graph theoretic manner using so-called trellis representations: see e.g. \cite{forney}. However, especially over large finite fields it is hard to algebraically describe a decoding algorithm and hence, a state space description as above is preferred. 

\end{remark}

The following theorems show how properties of a linear system are related to properties of the corresponding convolutional code.

\begin{theorem}\cite{bch}\ \\
$(A,B,C,D)$ is a minimal representation of $\mathcal{C}(A,B,C,D)$ if and only if it is reachable.
\end{theorem}

\begin{theorem}\cite{bch}\label{cor}\ \\
Assume that $(A,B,C,D)$ is reachable. Then $\mathcal{C}(A,B,C,D)$ is non-catastrophic if and only if $(A,B,C,D)$ is observable.
\end{theorem}

\section{Low-delay erasure decoding algorithm using the linear systems representation}

In this chapter, we develop our erasure decoding algorithm based on the ISO representation of the convolutional code. Some first ideas on decoding via this representation can already be found in \cite{vt}. We adopt some of the ideas presented there and combine it with new ideas to obtain a complete decoding algorithm.\\

Assume that we have a message $M=[m_0^{\top}\ \cdots\ m_{\gamma}^{\top}]^{\top}\in\mathbb F^{k(\gamma+1)}$ with $m_i\in\mathbb F^k$ which is sent at time step $i$. 
We write this message as $m(z)=\sum_{i=0}^{\gamma}m_{\gamma-i}z^i$ and encode it via a full rank, left prime, column reduced polynomial generator matrix $G(z)=\sum_{i=0}^{\mu}G_{\mu-i}z^i\in\mathbb F[z]^{n\times k}$ to obtain $v(z)=G(z)m(z)\in\mathbb F[z]^n$. We write $v(z)=\begin{pmatrix} y(z)\\ u(z) \end{pmatrix}$ with $y(z)=\sum_{i=0}^{\mu+\gamma}y_{\mu+\gamma-i}z^i\in\mathbb F[z]^{n-k}$ and $u(z)=\sum_{i=0}^{\mu+\gamma}u_{\mu+\gamma-i}z^i\in\mathbb F[z]^{k}$. As $m_0$ is sent first, we first receive $\begin{pmatrix} y_0\\ u_0 \end{pmatrix}=G_0m_0$, in the next time step $\begin{pmatrix} y_1\\ u_1 \end{pmatrix}=G_1m_0+G_0m_1$, and so on.

\begin{remark}
\normalfont
In principle, it would be also possible to encode the message via the linear system, i.e. to set $u(z)=m(z)$. In this case, one gets a rational generator matrix, which equals the transfer function of the linear system. But to make sure that the state and the output of the linear system have finite support, we had to impose restrictions on the input, i.e. on the message. This is why we consider this option as not suitable.
\end{remark}

Let $(A,B,C,D)$ be the linear systems representation of the convolutional code generated by $G(z)$.
Then, $(y_0, u_0, \hdots, y_j, u_j)$ represents the beginning of a codeword if and only if

\begin{align}\label{meq}
&\left[-I\ \vline\begin{array}{cccc}D & 0 & \hdots & 0\\ CB & \ddots & \ddots & \vdots\\ \vdots & \ddots & \ddots & 0\\CA^{j-1}B & \hdots & CB & D \end{array}\right]\begin{pmatrix}
y_0\\ \vdots\\ y_j\\ \hline u_0\\ \vdots\\ u_j
\end{pmatrix}\nonumber\\
&=\left[\begin{array}{ccccccc}-I & D & &  &  & 0\\ 0 & CB &  &  &  & \\ \vdots & & & \ddots & \ddots & \\0 & CA^{j-1}B  & \hdots & 0 & CB & -I & D \end{array}\right]\begin{pmatrix}
y_0\\u_0\\ \vdots\\ y_j\\ u_j
\end{pmatrix}=0
\end{align}

Moreover, one has for $i,j,l\in\mathbb N_0$:

\begin{align}\label{x}
\begin{pmatrix}
C\\ \vdots\\ CA^j
\end{pmatrix}x_{i+l}+\left[-I\ \vline\begin{array}{cccc}D & 0 & \hdots & 0\\ CB & \ddots & \ddots & \vdots\\ \vdots & \ddots & \ddots & 0\\CA^{j-1}B & \hdots & CB & D \end{array}\right]\begin{pmatrix}
y_{i+l}\\ \vdots\\ y_{i+l+j}\\ \hline u_{i+l}\\ \vdots\\ u_{i+l+j}
\end{pmatrix}=0
\end{align}
where
\begin{align}\label{b}
x_{i+l}=A^{i+l-1}Bu_0+\cdots+Bu_{i+l-1}.
\end{align}

Define $\mathcal{F}_0:=D$ and $\mathcal{F}_j:=\left[\begin{array}{cccc}D & 0 & \hdots & 0\\ CB & \ddots & \ddots & \vdots\\ \vdots & \ddots & \ddots & 0\\CA^{j-1}B & \hdots & CB & D \end{array}\right]$ for $j\geq 1$ as well as $\mathcal{R}_l:=[A^{l-1}B\cdots B]$ and $\ell:=\max\{l\ |\ \mathcal{R}_l\ \text{has full column rank}\}$ if $B$ has full column rank and $\ell:=-1$ otherwise.

\begin{theorem}\cite{mdp}
The quadruple $(A,B,C,D)$ is the linear systems representation of an MDP convolutional code if and only if each minor of $\mathcal{F}_L$ which is not
trivially zero is nonzero.
\end{theorem}

Furthermore, $u_i=y_i=0$ for $i>\gamma+\mu$ implies $$CA^{\gamma+\mu+w}Bu_0+\cdots+CA^wBu_{\gamma+\mu}=0$$ for $w\in\mathbb N_0$. Define $E_w:=\left[\begin{array}{ccc}CA^{\gamma+\mu}B &\cdots & CB\\ \vdots & & \vdots \\CA^{\gamma+\mu+w}B &\cdots & CA^wB\end{array}\right]$ and 
$\tilde{E}_w$ as submatrix of $E_w$ consisting only of the columns corresponding to components of $(u_0^\top,\hdots,u_{\gamma+\mu}^\top)$ that are not known yet.

We assume that the erasure recovering process has to be done within time delay $T$, i.e. it is neceassary that $m_i$ can be recovered after one has received (with possible erasures) $v_0, \hdots, v_i, \hdots, v_{i+T}$.

Assume that $v_0, \hdots, v_{i-1}$ are known and $v_i$ contains erasures. Then, one obtains
\begin{align}\label{recs}
&\left[-I\ \vline\begin{array}{cccc}D & 0 & \hdots & 0\\ CB & \ddots & \ddots & \vdots\\ \vdots & \ddots & \ddots & 0\\CA^{j-1}B & \hdots & CB & D \end{array}\right]\begin{pmatrix}
y_i\\ \vdots\\ y_{i+j}\\ \hline u_i\\ \vdots\\ u_{i+j}
\end{pmatrix}=\beta
\end{align}
where $\beta$ is a known vector depending on $v_0, \hdots, v_{i-1}$.\\

\textbf{Decoding Algorithm}\\
\textbf{1}: Set $i=-1$.\\
\textbf{2}: If there exists $w\in\mathbb N_0$ such that $\tilde{E}_w$ has full column rank,
go to 12, otherwise if $v_i$ contains erasures, go to 3 and if $v_i$ contains no erasures, set $i=i+1$ and repeat step 2.\\
\textbf{3}: Set $j=0$.\\
\textbf{4}: 
If $v_i$ can be recovered solving the linear system of equations induced by $[-I\ |\ \mathcal{F}_j]$ and $v_i,\hdots,v_{i+j}$ (see \eqref{recs}), go to 5, otherwise go to 6.\\
\textbf{5}: Recover the erasures in $v_i$ (and if possible also erasures in $v_{i+1},\hdots, v_{i+j}$), solving the system of linear equations \eqref{recs}. Replace the erased symbols with the correct symbols and go back to 2.\\
\textbf{6}: If $j=T$, we go to 7. Otherwise, we set $j=j+1$ and go back to 4.\\
\textbf{7}: Set $l=1$.\\
\textbf{8}: Set $j=0$.\\
\textbf{9}: If $x_{i+l}$ can be recovered solving the linear system of equations induced by \eqref{x} with $x_{i+l}$ and the erased components of $v_{i+l},\hdots, v_{i+l+j}$ as unknowns, we go to 10. Otherwise, we go to 11.\\
\textbf{10}: Recover $x_{i+l}$ and as much as possible of $v_{i+l},\hdots, v_{i+l+j}$ with the help of \eqref{x}. With the knowledge of $x_{i+l}$ and $u_0,\hdots,u_{i-1}$ and with equation \eqref{x}, obtain $A^{l-1}Bu_i+\cdots+Bu_{i+l-1}$. If $l\leq \ell$, this equations allows us to recover $u_i,\hdots,u_{i+l-1}$ and use it to compute $y_i,\hdots, y_{i+l-1}$ as well. If $l>\ell$ some values of $v_i,\hdots, v_{i+l-1}$ are lost but still we can restart the recovering process after these lost symbols. In either case, set $i=i+l-1$ and go back to 2.\\
\textbf{11}: If $j=T-l$, set $l=l+1$, and go back to 8. Otherwise set $j=j+1$ and go back to 9.\\
\textbf{12}: Use the system of linear equations $E_W\cdot[u_0^\top,\hdots,u_{\gamma+\mu}^\top]^\top=0$ to recover all erased components of $[u_0^\top,\hdots,u_{\gamma+\mu}^\top]^\top$. Afterwards use \eqref{meq} to obtain $[y_0^\top,\hdots,y_{\gamma+\mu}^\top]^\top$.\\

In steps 4 to 6 the algorithm recovers erasures forward within time delay $T$ as long as this is possible. If it reaches a point where this is not possible, it tries to recover the state of the corresponding linear system (steps 9 to 11) to be able to restart the decoding process (and recovers also symbols that had been lost in between, in case this is possible, even if these symbols are then recovered with a delay that is larger than $T$). After every successful recovery, in step 2, it is checked if there are already enough symbols known to recover the whole message with step 12. Note that due to theorem of Cayley-Hamilton one only has to check $\tilde{E}_w$ up to $w=\delta-1$.\\

In order to have a good performance for our algorithm, a convolutional code should fulfill the following properties as good as possible:

\begin{enumerate}
\item The nontrivial minors of $\mathcal{F}_j$ are nonzero for $j=1,\hdots,T$.
\item The nontrivial minors of $\left[\begin{matrix}
C & \vline\\ \vdots & \vline & \mathcal{F}_j\\ CA^j &
\vline \end{matrix}\right]$ are nonzero for $j=1,\hdots,T$.
\item For as many sets of columns of $E_w$ as possible, there exists $w=1,\hdots,\delta-1$ such that these columns are linearly independent.
\item $\ell$ is as large as possible.
\end{enumerate}

It is difficult to ensure that all these four properties are perfectly fulfilled. However, since these properties involve similar matrices, it seems to be a good attempt to construct a convolutional code in such a way that some of the properties are fulfilled, and then check how good the other properties are fulfilled. Clearly, if 2. is perfectly fulfilled, then also 1. Furthermore, there already exist constructions for matrices having all nontrivial minors nonzero (in the literature also referred to as superregular matrices); see e.g. \cite{dr13}, \cite{vt}, \cite{strongly}. Hence, to illustrate the performance of our algorithm with an example, we will construct a convolutional code such that 2. is perfectly fulfilled and then investigate how good 3. and 4. are fulfilled. Note that 4. is not so important for our algorithm as it only helps to recover symbols that had to be declared as lost with a larger delay as allowed by the delay constraint.\\ 

\begin{example}
\normalfont
We will construct an $(5,3,2)$ convolutional code for decoding with maximum delay $T=L=1$. First note that property 4 can never be fulfilled for these parameters because $\mathcal{R}_l$ has more columns than rows for all $l\in\mathbb N_0$. But as mentioned before, this property is only useful for the recovery of lost symbols with larger delay than originally prescribed and thus, it is no problem to neglect this. Hence, we want to construct $A, C\in\mathbb F^{2\times 2}$, $B, D\in\mathbb F^{2\times 3}$ such that $\left[\begin{matrix} C & D & 0 \\ CA & CB & D\end{matrix}\right]$ has all nontrivial minors nonzero for a suitable finite field $\mathbb F$. We use the construction for superregular matrices from \cite{dr16} as well as the fact that column permutation preserves superregularity to obtain that $$\left[\begin{matrix} C & D & 0 \\ CA & CB & D\end{matrix}\right]=\left[\begin{matrix} a^8 & a^{16} & a & a^2 & a^4 & 0 & 0 & 0 \\ a^{16} & a^{32} & a^2 & a^4 & a^8 & 0 & 0 & 0\\ a^{64} & a^{128} &  a^8 & a^{16} & a^{32} & a & a^2 & a^4\\a^{128} & a^{256} &  a^{16} & a^{32} & a^{64} & a^2 & a^4 & a^8\end{matrix}\right],$$ where $\mathbb F=\mathbb F_{p^N}$ with $N>330$ and $a$ is a primitive element of $\mathbb F$, has the property that all nontrivial minors are nonzero. We immediately obtain
$$D=\left[\begin{matrix}a & a^2 & a^4\\ a^2 & a^4 & a^8\end{matrix}\right]\quad\text{and}\quad C=\left[\begin{matrix} a^8 & a^{16}  \\ a^{16} & a^{32} \end{matrix}\right]$$
and can compute
$$B=C^{-1}(CB)=\left[\begin{matrix}1 & 0 & -a^{32}(a^8+1)\\ 0 & 1 & a^{16}(a^{16}+a^8+1)\end{matrix}\right]$$ and $$A=C^{-1}(CA)=\frac{1}{a^8-1}\left[\begin{matrix} a^{64}-a^{112} & a^{128}-a^{240}  \\ a^{104}-a^{48} & a^{232}-a^{112} \end{matrix}\right].$$

As $B$ is full rank, $(A,B,C,D)$ is a minimal ISO representation of an $(5,3,2)$ convolutional code $\mathcal{C}$ and since $\mathcal{F}_1$ is superregular, $\mathcal{C}$ is an MDP convolutional code. Hence, in particular, it has to fulfill Theorem \ref{cd} (ii), which is not possible if $G_1$ has two columns that are identically zero. Hence a generator matrix $G$ of $\mathcal{C}$ has at most one column degree that is equal to zero. Consequently, $G$ has column degrees $1,1,0$ since we assumed it to be a column reduced generator matrix and thus, the column degrees of $G$ have to sum up to $\delta=2$. Therefore, we obtain $\mu=1$.

Assume $\gamma=3$ and that we receive the following:\\

\begin{tabular}{|c| c| c|c|c|c|c|c|c|c|}
\hline
$y_0$ & $u_0$ &  $y_1$ & $u_1$ & $y_2$ & $u_2$ &  $y_3$ & $u_3$ & $y_4$ & $u_4$\\
\hline
$\ast$\ $\ast$\  &  $\surd$\ $\surd$ $\surd$ & $\ast$\ $\ast$\  &$\ast$ $\surd$\ $\surd$ & $\ast$\ $\ast$\  &  $\surd$\ $\surd$ $\surd$ & $\surd$\ $\surd$ & $\surd$  $\surd$ $\surd$ & $\ast$\ $\ast$\  & $\ast$\ $\ast$\  $\ast$\\
\hline
\end{tabular}\\

where $\ast$ symbolizes an erasure and $\surd$ a received symbol.

Since $\mathcal{C}$ is MDP, it can recover $n-k$ erasures out of $n$ symbols or $2(n-k)$ erasures out of $2n$ symbols (assuming that there are no erasures in front of this window of size $n$ or $2n$, respectively). The steps of our algorithm with $\mathcal{C}$ and the above erasure pattern would be the following.

First, the algorithm uses \eqref{recs} with $j=0$ to recover $y_0$. Afterwards, one realizes that it is neither possible to recover $y_1$ and $u_1$ with \eqref{recs} for $j=0$ nor $y_1, u_1, y_2, u_2$ with \eqref{recs} for $j=1$. The algorithm applies \eqref{x} with $i=l=1$ to recover $x_2$ and $y_2$ but the erased components of $y_1$ and $u_1$ have to be declared as lost. Finally, as the matrix consisting of the first column of $\begin{pmatrix}
CA^3B\\CA^4B
\end{pmatrix}$ and all columns of $\begin{pmatrix}
CB\\CAB
\end{pmatrix}$ has nonzero determinant, one can use step 12 of the algorithm to recover the lost component of $u_1$ as well as $u_4$ before $u_4$ and $y_4$ were even sent, just with the knowledge of the already known symbols of $u_0, u_1, u_2, u_3$ and with the information that $\gamma=3$, i.e. $u_i=y_i=0$ for $i>4$. Then, with the knowledge of $u_0,\hdots,u_4$, it is also possible to compute the erased components of $y_1$ and $y_4$. In summary, we are able to recover the whole sequence but part of it only with a larger delay than actually allowed. However, we were able to obtain $u_4, y_4$ already one time interval before these vectors were sent, i.e. in some sense with delay $-1$.
\end{example}

\section{Performance Analysis}

In this section, we will explain the two main advantages of our systems theoretic decoding algorithm with respect to the (first) erasure decoding algorithm for convolutional codes that can be found in \cite{vp}, namely the reduced decoding delay and the reduced computational effort.

Our algorithm tries to recover the occurring erasures with smallest possible delay by first trying to do the recovery in a window of size $n$, afterwards in a window of size $2n$, and so on. In contrast to this approach, the decoding algorithm in \cite{vp} first tries to decode in the largest possible window of size $(L+1)n$ and only decreases this window if it fails to recover all the erasures in the big window. This implies that the decoding delay is always at least $L$. Moreover, it is computationally less complex and less costly to do several decoding steps in small windows than one decoding step in a larger window whose size is the sum of the sizes of the smaller windows since it is easier to solve several small than one large linear system of equations. In addition, by using the linear systems approach, the systems of equations we have to solve for erasure recovery are parts of linear systems that are already in echelon form; see \eqref{recs}. Especially, when we transmit over a channel with a statistic that implies that it is more likely to get erasures in the $y_i$ than in the $u_i$, this is of very big advantage as you can obtain any erased component of any $y_i$ (that has the possibility to be recovered), directly from \eqref{recs} with very small computational effort.

Finally, as we already observed in our example, the use of the terminating equations in step 12 of the algorithm can make it possile to obtain symbols that were not even sent yet, i.e. in some sense we are able to "look into the future" and terminate the decoding before the end of the transmission. This is of course an additional considerable reduction of the decoding delay.

\section{Conclusion}
In this paper, we presented an erasure decoding algorithm for convolutional codes employing their linear systems representation. We observed that this algorithm is able to reduce the decoding delay and the computational effort in comparison with previous algorithms.

\section*{Acknowledgments}
The authors acknowledge the support of Swiss National Science Foundation grant n. 188430.  Julia Lieb acknowledges also the support of the German Research Foundation grant LI 3101/1-1.

\bibliography{mybibfile}

\end{document}